\documentclass[conference,10pt,twocolumn,letter]{IEEEtran}
\IEEEoverridecommandlockouts
\usepackage{cite}
\usepackage[T1]{fontenc}
\usepackage{graphicx}

\usepackage{amssymb}
\usepackage{amsmath}
\usepackage{paralist}
\usepackage{subfigure}
\usepackage{booktabs} 
\usepackage{algorithm}
\usepackage{algorithmic}
\usepackage{amsthm}
\usepackage{multirow}
\usepackage{microtype} 
\usepackage{tablefootnote}
\usepackage{balance}
\usepackage[paper=letterpaper,top=0.7in,bottom=1.0in,right=0.68in,left=0.68in]{geometry}

\usepackage{amssymb}
\usepackage{amsfonts}
\usepackage{mathrsfs}
\usepackage{xspace}
\usepackage{bm}
\usepackage{upgreek}

\newcommand{\safemath}[2]{\newcommand{#1}{\ensuremath{#2}\xspace}}

\safemath{\bma}{\mathbf{a}}
\safemath{\bmb}{\mathbf{b}}
\safemath{\bmc}{\mathbf{c}}
\safemath{\bmd}{\mathbf{d}}
\safemath{\bme}{\mathbf{e}}
\safemath{\bmf}{\mathbf{f}}
\safemath{\bmg}{\mathbf{g}}
\safemath{\bmh}{\mathbf{h}}
\safemath{\bmi}{\mathbf{i}}
\safemath{\bmj}{\mathbf{j}}
\safemath{\bmk}{\mathbf{k}}
\safemath{\bml}{\mathbf{l}}
\safemath{\bmm}{\mathbf{m}}
\safemath{\bmn}{\mathbf{n}}
\safemath{\bmo}{\mathbf{o}}
\safemath{\bmp}{\mathbf{p}}
\safemath{\bmq}{\mathbf{q}}
\safemath{\bmr}{\mathbf{r}}
\safemath{\bms}{\mathbf{s}}
\safemath{\bmt}{\mathbf{t}}
\safemath{\bmu}{\mathbf{u}}
\safemath{\bmv}{\mathbf{v}}
\safemath{\bmw}{\mathbf{w}}
\safemath{\bmx}{\mathbf{x}}
\safemath{\bmy}{\mathbf{y}}
\safemath{\bmz}{\mathbf{z}}
\safemath{\bmzero}{\mathbf{0}}
\safemath{\bmone}{\mathbf{1}}

\bmdefine{\biad}{a}
\bmdefine{\bibd}{b}
\bmdefine{\bicd}{c}
\bmdefine{\bidd}{d}
\bmdefine{\bied}{e}
\bmdefine{\bifd}{f}
\bmdefine{\bigd}{g}
\bmdefine{\bihd}{h}
\bmdefine{\biid}{i}
\bmdefine{\bijd}{j}
\bmdefine{\bikd}{k}
\bmdefine{\bild}{l}
\bmdefine{\bimd}{m}
\bmdefine{\bind}{n}
\bmdefine{\biod}{o}
\bmdefine{\bipd}{p}
\bmdefine{\biqd}{q}
\bmdefine{\bird}{r}
\bmdefine{\bisd}{s}
\bmdefine{\bitd}{t}
\bmdefine{\biud}{u}
\bmdefine{\bivd}{v}
\bmdefine{\biwd}{w}
\bmdefine{\bixd}{x}
\bmdefine{\biyd}{y}
\bmdefine{\bizd}{z}

\bmdefine{\bixid}{\xi}
\bmdefine{\bilambdad}{\lambda}
\bmdefine{\bimud}{\mu}
\bmdefine{\bithetad}{\theta}
\bmdefine{\biphid}{\phi}
\bmdefine{\bideltad}{\delta}

\safemath{\bmia}{\biad}
\safemath{\bmib}{\bibd}
\safemath{\bmic}{\bicd}
\safemath{\bmid}{\bidd}
\safemath{\bmie}{\bied}
\safemath{\bmif}{\bifd}
\safemath{\bmig}{\bigd}
\safemath{\bmih}{\bihd}
\safemath{\bmii}{\biid}
\safemath{\bmij}{\bijd}
\safemath{\bmik}{\bikd}
\safemath{\bmil}{\bild}
\safemath{\bmim}{\bimd}
\safemath{\bmin}{\bind}
\safemath{\bmio}{\biod}
\safemath{\bmip}{\bipd}
\safemath{\bmiq}{\biqd}
\safemath{\bmir}{\bird}
\safemath{\bmis}{\bisd}
\safemath{\bmit}{\bitd}
\safemath{\bmiu}{\biud}
\safemath{\bmiv}{\bivd}
\safemath{\bmiw}{\biwd}
\safemath{\bmix}{\bixd}
\safemath{\bmiy}{\biyd}
\safemath{\bmiz}{\bizd}

\safemath{\bmxi}{\bixid}
\safemath{\bmlambda}{\bilambdad}
\safemath{\bmmu}{\bimud}
\safemath{\bmtheta}{\bithetad}
\safemath{\bmphi}{\biphid}
\safemath{\bmdelta}{\bideltad}

\safemath{\bA}{\mathbf{A}}
\safemath{\bB}{\mathbf{B}}
\safemath{\bC}{\mathbf{C}}
\safemath{\bD}{\mathbf{D}}
\safemath{\bE}{\mathbf{E}}
\safemath{\bF}{\mathbf{F}}
\safemath{\bG}{\mathbf{G}}
\safemath{\bH}{\mathbf{H}}
\safemath{\bI}{\mathbf{I}}
\safemath{\bJ}{\mathbf{J}}
\safemath{\bK}{\mathbf{K}}
\safemath{\bL}{\mathbf{L}}
\safemath{\bM}{\mathbf{M}}
\safemath{\bN}{\mathbf{N}}
\safemath{\bO}{\mathbf{O}}
\safemath{\bP}{\mathbf{P}}
\safemath{\bQ}{\mathbf{Q}}
\safemath{\bR}{\mathbf{R}}
\safemath{\bS}{\mathbf{S}}
\safemath{\bT}{\mathbf{T}}
\safemath{\bU}{\mathbf{U}}
\safemath{\bV}{\mathbf{V}}
\safemath{\bW}{\mathbf{W}}
\safemath{\bX}{\mathbf{X}}
\safemath{\bY}{\mathbf{Y}}
\safemath{\bZ}{\mathbf{Z}}

\safemath{\bZero}{\mathbf{0}}
\safemath{\bOne}{\mathbf{1}}
\safemath{\bDelta}{\mathbf{\Delta}}
\safemath{\bLambda}{\mathbf{\UpLambda}}
\safemath{\bPhi}{\mathbf{\Upphi}}
\safemath{\bSigma}{\mathbf{\Upsigma}}
\safemath{\bOmega}{\mathbf{\Upomega}}
\safemath{\bTheta}{\mathbf{\Uptheta}}

\bmdefine{\biAd}{A}
\bmdefine{\biBd}{B}
\bmdefine{\biCd}{C}
\bmdefine{\biDd}{D}
\bmdefine{\biEd}{E}
\bmdefine{\biFd}{F}
\bmdefine{\biGd}{G}
\bmdefine{\biHd}{H}
\bmdefine{\biId}{I}
\bmdefine{\biJd}{J}
\bmdefine{\biKd}{K}
\bmdefine{\biLd}{L}
\bmdefine{\biMd}{M}
\bmdefine{\biOd}{N}
\bmdefine{\biPd}{O}
\bmdefine{\biQd}{P}
\bmdefine{\biRd}{R}
\bmdefine{\biSd}{S}
\bmdefine{\biTd}{T}
\bmdefine{\biUd}{U}
\bmdefine{\biVd}{V}
\bmdefine{\biWd}{W}
\bmdefine{\biXd}{X}
\bmdefine{\biYd}{Y}
\bmdefine{\biZd}{Z}

\bmdefine{\biDelta}{\Delta}
\bmdefine{\biLambda}{\Lambda}
\bmdefine{\biPhi}{\Phi}
\bmdefine{\biSigma}{\Sigma}
\bmdefine{\biOmega}{\Omega}
\bmdefine{\biTheta}{\Theta}

\safemath{\bimA}{\biAd}
\safemath{\bimB}{\biBd}
\safemath{\bimC}{\biCd}
\safemath{\bimD}{\biDd}
\safemath{\bimE}{\biEd}
\safemath{\bimF}{\biFd}
\safemath{\bimG}{\biGd}
\safemath{\bimH}{\biHd}
\safemath{\bimI}{\biId}
\safemath{\bimJ}{\biJd}
\safemath{\bimK}{\biKd}
\safemath{\bimL}{\biLd}
\safemath{\bimM}{\biMd}
\safemath{\bimN}{\biNd}
\safemath{\bimO}{\biOd}
\safemath{\bimP}{\biPd}
\safemath{\bimQ}{\biQd}
\safemath{\bimR}{\biRd}
\safemath{\bimS}{\biSd}
\safemath{\bimT}{\biTd}
\safemath{\bimU}{\biUd}
\safemath{\bimV}{\biVd}
\safemath{\bimW}{\biWd}
\safemath{\bimX}{\biXd}
\safemath{\bimY}{\biYd}
\safemath{\bimZ}{\biZd}

\safemath{\bimDelta}{\biDelta}
\safemath{\bimLambda}{\biLambda}
\safemath{\bimPhi}{\biPhi}
\safemath{\bimSigma}{\biSigma}
\safemath{\bimOmega}{\biOmega}
\safemath{\bimTheta}{\biTheta}

\safemath{\setA}{\mathcal{A}}
\safemath{\setB}{\mathcal{B}}
\safemath{\setC}{\mathcal{C}}
\safemath{\setD}{\mathcal{D}}
\safemath{\setE}{\mathcal{E}}
\safemath{\setF}{\mathcal{F}}
\safemath{\setG}{\mathcal{G}}
\safemath{\setH}{\mathcal{H}}
\safemath{\setI}{\mathcal{I}}
\safemath{\setJ}{\mathcal{J}}
\safemath{\setK}{\mathcal{K}}
\safemath{\setL}{\mathcal{L}}
\safemath{\setM}{\mathcal{M}}
\safemath{\setN}{\mathcal{N}}
\safemath{\setO}{\mathcal{O}}
\safemath{\setP}{\mathcal{P}}
\safemath{\setQ}{\mathcal{Q}}
\safemath{\setR}{\mathcal{R}}
\safemath{\setS}{\mathcal{S}}
\safemath{\setT}{\mathcal{T}}
\safemath{\setU}{\mathcal{U}}
\safemath{\setV}{\mathcal{V}}
\safemath{\setW}{\mathcal{W}}
\safemath{\setX}{\mathcal{X}}
\safemath{\setY}{\mathcal{Y}}
\safemath{\setZ}{\mathcal{Z}}
\safemath{\emptySet}{\varnothing}

\safemath{\colA}{\mathscr{A}}
\safemath{\colB}{\mathscr{B}}
\safemath{\colC}{\mathscr{C}}
\safemath{\colD}{\mathscr{D}}
\safemath{\colE}{\mathscr{E}}
\safemath{\colF}{\mathscr{F}}
\safemath{\colG}{\mathscr{G}}
\safemath{\colH}{\mathscr{H}}
\safemath{\colI}{\mathscr{I}}
\safemath{\colJ}{\mathscr{J}}
\safemath{\colK}{\mathscr{K}}
\safemath{\colL}{\mathscr{L}}
\safemath{\colM}{\mathscr{M}}
\safemath{\colN}{\mathscr{N}}
\safemath{\colO}{\mathscr{O}}
\safemath{\colP}{\mathscr{P}}
\safemath{\colQ}{\mathscr{Q}}
\safemath{\colR}{\mathscr{R}}
\safemath{\colS}{\mathscr{S}}
\safemath{\colT}{\mathscr{T}}
\safemath{\colU}{\mathscr{U}}
\safemath{\colV}{\mathscr{V}}
\safemath{\colW}{\mathscr{W}}
\safemath{\colX}{\mathscr{X}}
\safemath{\colY}{\mathscr{Y}}
\safemath{\colZ}{\mathscr{Z}}

\safemath{\opA}{\mathbb{A}}
\safemath{\opB}{\mathbb{B}}
\safemath{\opC}{\mathbb{C}}
\safemath{\opD}{\mathbb{D}}
\safemath{\opE}{\mathbb{E}}
\safemath{\opF}{\mathbb{F}}
\safemath{\opG}{\mathbb{G}}
\safemath{\opH}{\mathbb{H}}
\safemath{\opI}{\mathbb{I}}
\safemath{\opJ}{\mathbb{J}}
\safemath{\opK}{\mathbb{K}}
\safemath{\opL}{\mathbb{L}}
\safemath{\opM}{\mathbb{M}}
\safemath{\opN}{\mathbb{N}}
\safemath{\opO}{\mathbb{O}}
\safemath{\opP}{\mathbb{P}}
\safemath{\opQ}{\mathbb{Q}}
\safemath{\opR}{\mathbb{R}}
\safemath{\opS}{\mathbb{S}}
\safemath{\opT}{\mathbb{T}}
\safemath{\opU}{\mathbb{U}}
\safemath{\opV}{\mathbb{V}}
\safemath{\opW}{\mathbb{W}}
\safemath{\opX}{\mathbb{X}}
\safemath{\opY}{\mathbb{Y}}
\safemath{\opZ}{\mathbb{Z}}
\safemath{\opZero}{\mathbb{O}}
\safemath{\identityop}{\opI}

\safemath{\veca}{\bma}
\safemath{\vecb}{\bmb}
\safemath{\vecc}{\bmc}
\safemath{\vecd}{\bmd}
\safemath{\vece}{\bme}
\safemath{\vecf}{\bmf}
\safemath{\vecg}{\bmg}
\safemath{\vech}{\bmh}
\safemath{\veci}{\bmi}
\safemath{\vecj}{\bmj}
\safemath{\veck}{\bmk}
\safemath{\vecl}{\bml}
\safemath{\vecm}{\bmm}
\safemath{\vecn}{\bmn}
\safemath{\veco}{\bmo}
\safemath{\vecp}{\bmp}
\safemath{\vecq}{\bmq}
\safemath{\vecr}{\bmr}
\safemath{\vecs}{\bms}
\safemath{\vect}{\bmt}
\safemath{\vecu}{\bmu}
\safemath{\vecv}{\bmv}
\safemath{\vecw}{\bmw}
\safemath{\vecx}{\bmx}
\safemath{\vecy}{\bmy}
\safemath{\vecz}{\bmz}

\safemath{\veczero}{\bmzero}
\safemath{\vecone}{\bmone}
\safemath{\vecxi}{\bmxi}
\safemath{\veclambda}{\bmlambda}
\safemath{\vecmu}{\bmmu}
\safemath{\vectheta}{\bmtheta}
\safemath{\vecphi}{\bmphi}
\safemath{\vecdelta}{\bmdelta}

\safemath{\matA}{\bA}
\safemath{\matB}{\bB}
\safemath{\matC}{\bC}
\safemath{\matD}{\bD}
\safemath{\matE}{\bE}
\safemath{\matF}{\bF}
\safemath{\matG}{\bG}
\safemath{\matH}{\bH}
\safemath{\matI}{\bI}
\safemath{\matJ}{\bJ}
\safemath{\matK}{\bK}
\safemath{\matL}{\bL}
\safemath{\matM}{\bM}
\safemath{\matN}{\bN}
\safemath{\matO}{\bO}
\safemath{\matP}{\bP}
\safemath{\matQ}{\bQ}
\safemath{\matR}{\bR}
\safemath{\matS}{\bS}
\safemath{\matT}{\bT}
\safemath{\matU}{\bU}
\safemath{\matV}{\bV}
\safemath{\matW}{\bW}
\safemath{\matX}{\bX}
\safemath{\matY}{\bY}
\safemath{\matZ}{\bZ}
\safemath{\matzero}{\bmzero}

\safemath{\matDelta}{\bDelta}
\safemath{\matLambda}{\bLambda}
\safemath{\matPhi}{\bPhi}
\safemath{\matSigma}{\bSigma}
\safemath{\matOmega}{\bOmega}
\safemath{\matTheta}{\bTheta}

\safemath{\matidentity}{\matI}
\safemath{\matone}{\matO}

\safemath{\rnda}{A}
\safemath{\rndb}{B}
\safemath{\rndc}{C}
\safemath{\rndd}{D}
\safemath{\rnde}{E}
\safemath{\rndf}{F}
\safemath{\rndg}{G}
\safemath{\rndh}{H}
\safemath{\rndi}{I}
\safemath{\rndj}{J}
\safemath{\rndk}{K}
\safemath{\rndl}{L}
\safemath{\rndm}{M}
\safemath{\rndn}{N}
\safemath{\rndo}{O}
\safemath{\rndp}{P}
\safemath{\rndq}{Q}
\safemath{\rndr}{R}
\safemath{\rnds}{S}
\safemath{\rndt}{T}
\safemath{\rndu}{U}
\safemath{\rndv}{V}
\safemath{\rndw}{W}
\safemath{\rndx}{X}
\safemath{\rndy}{Y}
\safemath{\rndz}{Z}

\safemath{\rveca}{\bimA}
\safemath{\rvecb}{\bimB}
\safemath{\rvecc}{\bimC}
\safemath{\rvecd}{\bimD}
\safemath{\rvece}{\bimE}
\safemath{\rvecf}{\bimF}
\safemath{\rvecg}{\bimG}
\safemath{\rvech}{\bimH}
\safemath{\rveci}{\bimI}
\safemath{\rvecj}{\bimJ}
\safemath{\rveck}{\bimK}
\safemath{\rvecl}{\bimL}
\safemath{\rvecm}{\bimM}
\safemath{\rvecn}{\bimN}
\safemath{\rveco}{\bomO}
\safemath{\rvecp}{\bimP}
\safemath{\rvecq}{\bimQ}
\safemath{\rvecr}{\bimR}
\safemath{\rvecs}{\bimS}
\safemath{\rvect}{\bimT}
\safemath{\rvecu}{\bimU}
\safemath{\rvecv}{\bimV}
\safemath{\rvecw}{\bimW}
\safemath{\rvecx}{\bimX}
\safemath{\rvecy}{\bimY}
\safemath{\rvecz}{\bimZ}

\safemath{\rvecxi}{\bmxi}
\safemath{\rveclambda}{\bmlambda}
\safemath{\rvecmu}{\bmmu}
\safemath{\rvectheta}{\bmtheta}
\safemath{\rvecphi}{\bmphi}

\safemath{\rmatA}{\bimA}
\safemath{\rmatB}{\bimB}
\safemath{\rmatC}{\bimC}
\safemath{\rmatD}{\bimD}
\safemath{\rmatE}{\bimE}
\safemath{\rmatF}{\bimF}
\safemath{\rmatG}{\bimG}
\safemath{\rmatH}{\bimH}
\safemath{\rmatI}{\bimI}
\safemath{\rmatJ}{\bimJ}
\safemath{\rmatK}{\bimK}
\safemath{\rmatL}{\bimL}
\safemath{\rmatM}{\bimM}
\safemath{\rmatN}{\bimN}
\safemath{\rmatO}{\bimO}
\safemath{\rmatP}{\bimP}
\safemath{\rmatQ}{\bimQ}
\safemath{\rmatR}{\bimR}
\safemath{\rmatS}{\bimS}
\safemath{\rmatT}{\bimT}
\safemath{\rmatU}{\bimU}
\safemath{\rmatV}{\bimV}
\safemath{\rmatW}{\bimW}
\safemath{\rmatX}{\bimX}
\safemath{\rmatY}{\bimY}
\safemath{\rmatZ}{\bimZ}

\safemath{\rmatDelta}{\bimDelta}
\safemath{\rmatLambda}{\bimLambda}
\safemath{\rmatPhi}{\bimPhi}
\safemath{\rmatSigma}{\bimSigma}
\safemath{\rmatOmega}{\bimOmega}
\safemath{\rmatTheta}{\bimTheta}

\usepackage{amssymb}
\usepackage{amsfonts}
\usepackage{mathrsfs}
\usepackage{xspace}
\usepackage{bm}
\usepackage{fancyref}
\usepackage{textcomp}

\usepackage{multirow}
\usepackage{stmaryrd}

\newenvironment{textbmatrix}{	\setlength{\arraycolsep}{2.5pt}%
								\big[\begin{matrix}}{\end{matrix}\big]%
								\raisebox{0.08ex}{\vphantom{M}}}

\def\be{\begin{equation}}
\def\ee{\end{equation}}
\def\een{\nonumber \end{equation}}
\def\mat{\begin{bmatrix}}
\def\emat{\end{bmatrix}}
\def\btm{\begin{textbmatrix}}
\def\etm{\end{textbmatrix}}

\def\ba#1\ea{\begin{align}#1\end{align}}
\def\bas#1\eas{\begin{align*}#1\end{align*}}
\def\bs#1\es{\begin{split}#1\end{split}}
\def\bg#1\eg{\begin{gather}#1\end{gather}}
\def\bml#1\eml{\begin{multline}#1\end{multline}}
\def\bi#1\ei{\begin{itemize}#1\end{itemize}}

\DeclareMathOperator{\sign}{sign}			
\DeclareMathOperator*{\argmin}{arg\;min}		

\safemath{\dirac}{\delta}					
\safemath{\krond}{\dirac}					

\safemath{\upto}{\uparrow}
\safemath{\downto}{\downarrow}
\safemath{\iu}{j}							
\safemath{\ev}{\lambda}						
\safemath{\hilseqspace}{l^{2}}				
\newcommand{\banachfunspace}[1]{\setL^{#1}}	
\safemath{\hilfunspace}{\banachfunspace{2}}	

\safemath{\SNR}{\textit{SNR}} 				
\safemath{\PAR}{\textit{PAR}} 				
\safemath{\No}{N_0}							
\safemath{\Es}{E_s}							
\safemath{\Eb}{E_b}							
\safemath{\EbNo}{\frac{\Eb}{\No}}
\safemath{\EsNo}{\frac{\Es}{\No}}

\DeclareMathOperator{\CHop}{\ensuremath{\opH}} 
\safemath{\tvir}{\rndh_{\CHop}}				
\safemath{\tvtf}{\rndl_{\CHop}}				
\safemath{\spf}{\rnds_{\CHop}}				
\safemath{\bff}{H_{\CHop}}					

\safemath{\ircf}{r_{h}}						
\safemath{\tftvcf}{r_{s}}					
\safemath{\tfcf}{r_{l}}						
\safemath{\bfcf}{r_{H}}						

\safemath{\tcorr}{c_h}						
\safemath{\scf}{c_{s}}						
\safemath{\tfcorr}{c_{l}}					
\safemath{\fcorr}{c_{H}}						

\safemath{\mi}{I}							
\safemath{\capacity}{C}						

\safemath{\normal}{\mathcal{N}}			
\safemath{\jpg}{\mathcal{CN}}			
\safemath{\mchain}{\leftrightarrow}		

\safemath{\dB}{\,\mathrm{dB}}
\safemath{\dBm}{\,\mathrm{dBm}}
\safemath{\Hz}{\,\mathrm{Hz}}
\safemath{\kHz}{\,\mathrm{kHz}}
\safemath{\MHz}{\,\mathrm{MHz}}
\safemath{\GHz}{\,\mathrm{GHz}}
\safemath{\s}{\,\mathrm{s}}
\safemath{\ms}{\,\mathrm{ms}}
\safemath{\mus}{\,\mathrm{\text{\textmu}s}}
\safemath{\ns}{\,\mathrm{ns}}
\safemath{\ps}{\,\mathrm{ps}}
\safemath{\meter}{\,\mathrm{m}}
\safemath{\mm}{\,\mathrm{mm}}
\safemath{\cm}{\,\mathrm{cm}}
\safemath{\m}{\,\mathrm{m}}
\safemath{\W}{\,\mathrm{W}}
\safemath{\mW}{\, \mathrm{mW}}
\safemath{\J}{\,\mathrm{J}}
\safemath{\K}{\,\mathrm{K}}
\safemath{\bit}{\,\mathrm{bit}}
\safemath{\nat}{\,\mathrm{nat}}

\safemath{\define}{\triangleq}			

\safemath{\equivalent}{\sim}
\safemath{\distas}{\sim}					
\safemath{\sdiff}{\Delta}				

\safemath{\reals}{\mathbb{R}}
\safemath{\positivereals}{\reals_{+}}
\safemath{\integers}{\mathbb{Z}}
\safemath{\posint}{\integers_{+}}
\safemath{\naturals}{\mathbb{N}}
\safemath{\posnaturals}{\naturals_{+}}
\safemath{\complexset}{\mathbb{C}}
\safemath{\rationals}{\mathbb{Q}}

\newcommand*{\fancyrefapplabelprefix}{app}		
\newcommand*{\fancyrefthmlabelprefix}{thm}		
\newcommand*{\fancyreflemlabelprefix}{lem}		
\newcommand*{\fancyrefcorlabelprefix}{cor}		
\newcommand*{\fancyrefdeflabelprefix}{def}		
\newcommand*{\fancyrefproplabelprefix}{prop}		
\newcommand*{\fancyrefexmpllabelprefix}{exmpl}
\newcommand*{\fancyrefalglabelprefix}{alg}		
\newcommand*{\fancyreftbllabelprefix}{tbl}		

\frefformat{vario}{\fancyrefseclabelprefix}{Section~#1}
\frefformat{vario}{\fancyrefthmlabelprefix}{Theorem~#1}
\frefformat{vario}{\fancyreftbllabelprefix}{Table~#1}
\frefformat{vario}{\fancyreflemlabelprefix}{Lemma~#1}
\frefformat{vario}{\fancyrefcorlabelprefix}{Corollary~#1}
\frefformat{vario}{\fancyrefdeflabelprefix}{Definition~#1}
\frefformat{vario}{\fancyreffiglabelprefix}{Fig.~#1}
\frefformat{vario}{\fancyrefapplabelprefix}{Appendix~#1}
\frefformat{vario}{\fancyrefeqlabelprefix}{(#1)}
\frefformat{vario}{\fancyrefproplabelprefix}{Proposition~#1}
\frefformat{vario}{\fancyrefexmpllabelprefix}{Example~#1}
\frefformat{vario}{\fancyrefalglabelprefix}{Algorithm~#1}

\usepackage{color}

\allowdisplaybreaks 

\newcommand{\rhat}[0]{\hat{\bmr}} 
\newcommand{\yhat}[0]{\hat{\bmy}} 
\newcommand{\rshat}[0]{\hat{\bmr}^s} 
\newcommand{\hhat}[0]{\hat{\bmh}} 
\newcommand{\dhat}[0]{\hat{\bmd}} 
\newcommand{\ehat}[0]{\hat{\bme}} 
\newcommand{\qhat}[0]{\hat{\bmq}} 
\newcommand{\hhatstar}[0]{\hat{\bmh}^\star} 
\newcommand{\hhattick}[0]{\hat{\bmh}^\prime}

\newcommand{\mur}[0]{\mu(\rhat)} 
\newcommand{\mure}[1]{\mu(\rhate{#1})}

\newcommand{\rhate}[1]{\hat{r}_{#1}} 
\newcommand{\yhate}[1]{\hat{y}_{#1}}
\newcommand{\rshate}[1]{\hat{r}^s_{#1}}

\newcommand{\hhatstare}[1]{\hat{h}^\star_{#1}}

\newcommand{\E}[1]{\mathbb{E}\!\left[#1\right]} 
\newcommand{\smolE}[1]{\mathbb{E}[#1]} 

\newcommand{\realindex}[1]{[#1]_{\mathcal{R}}}
\newcommand{\imagindex}[1]{[#1]_{\mathcal{I}}}
\newcommand{\realpart}[1]{\mathcal{R}\{#1\}}
\newcommand{\imagpart}[1]{\mathcal{I}\{#1\}}

\safemath{\SURE}{\textit{SURE}}
\safemath{\MSE}{\textit{MSE}}
\safemath{\EVM}{\textit{EVM}}

\safemath{\Eo}{E_0}
\safemath{\Eh}{E_h}
\safemath{\Ep}{E_p}
\safemath{\Do}{D_0}
\safemath{\Qo}{Q_0}
\safemath{\Dbeta}{D_\beta}
\safemath{\dalpha}{{\bmd}_{\alpha}}
\safemath{\dhatalpha}{\hat{\bmd}_{\alpha}}
\safemath{\dhatbeta}{\hat{\bmd}_{\beta}}

\safemath{\Tran}{\textnormal{T}}
\safemath{\Herm}{\textnormal{H}}

\safemath{\CN}{\mathcal{CN}}
\safemath{\N}{\mathcal{N}}

\safemath{\diag}{\textnormal{diag}}
\safemath{\trace}{\textnormal{trace}}

\safemath{\sumabsysquared}{a}
\safemath{\sumabsy}{b}
\safemath{\sumone}{(B-k)}
\safemath{\sumabsyinverse}{c}

\begin{document}

\title{Sparsity-Adaptive Beamspace Channel Estimation for 1-Bit mmWave Massive MIMO Systems}

\author{\IEEEauthorblockN{Alexandra Gallyas-Sanhueza$^\text{1}$, Seyed Hadi Mirfarshbafan$^\text{1}$, Ramina Ghods$^\text{2}$, and Christoph Studer$^\text{1}$} \\[-0.1cm]
\IEEEauthorblockA{
\textit{$^\text{1}$Cornell Tech, New York, NY;} \textit{$^\text{2}$Carnegie Mellon University, Pittsburgh, PA}
} 
\thanks{The work of RG and CS was supported in part by the US NSF under grants ECCS-1408006, CCF-1535897,  CCF-1652065, CNS-1717559, and ECCS-1824379. The work of AGS, SHM, and CS was supported in part by Xilinx, Inc.\ and by ComSenTer, one of six centers in JUMP, a Semiconductor Research Corporation (SRC) program sponsored by DARPA.  
}\\[-0.2cm]
}

\maketitle
\begin{abstract}
We propose sparsity-adaptive beamspace channel estimation algorithms that improve accuracy for 1-bit data converters in all-digital millimeter-wave (mmWave) massive multiple-input multiple-output (MIMO)  basestations. Our algorithms include a tuning stage based on Stein's unbiased risk estimate (SURE) that automatically selects optimal denoising parameters depending on the instantaneous channel conditions. Simulation results with line-of-sight (LoS) and non-LoS mmWave massive MIMO channel models show that our algorithms improve channel estimation accuracy with 1-bit measurements in a computationally-efficient manner.
\end{abstract}

\section{Introduction}
Millimeter-wave (mmWave) and massive multi-user (MU) multiple-input multiple-output (MIMO) will be  core technologies for future wireless systems~\cite{larsson14a, rappaport15a}.
The combination of these technologies enables simultaneous communication to multiple user equipments (UEs) at unprecedentedly high data rates. 
These advantages come at the cost of significantly increased power consumption, implementation complexity, and system costs. A viable solution to address these challenges is the use of low-resolution data converters combined with sophisticated but efficient baseband processing algorithms in all-digital basestations (BS) architectures~\cite{dutta2019case,jacobsson17b,li17b,mo16b,panagiotis20}. 

\subsection{Channel Estimation with Low-Resolution Data Converters}
Coarse quantization of the received baseband samples, due to the use of low-resolution analog-to-digital converters (ADCs) at the BS, together with the high path loss at mmWave or terahertz (THz) frequencies~\cite{rappaport15b, gao16}, renders the acquisition of accurate channel estimates a challenging task.
Fortunately,  wave propagation at mmWave or THz frequencies is directional~\cite{akdeniz14a} and channels typically consist only of a few dominant propagation paths~\cite{rappaport13a,rappaport15a}. Both of these properties cause the channel vectors to be sparse in the beamspace domain, which can be exploited to perform denoising that improves reliability of data transmission~\cite{alkhateeb14a,mo14b,tang13,brady13,ghods19a}. 

Practical sparsity-exploiting channel denoising methods for mmWave massive MU-MIMO systems must exhibit low computational complexity due to the large number of BS antenna elements and the potentially large number of UEs that commmunicate simultaneously.
A low-complexity mmWave channel denoising algorithm called BEACHES (short for beamspace channel estimation) has been proposed recently in~\cite{ghods19a}. This method has orders-of-magnitude lower complexity than state-of-the-art denoising methods, such as atomic norm minimization (ANM)~\cite{bhaskar13} and Newtonized orthogonal matching pursuit (NOMP) \cite{mamandipoor16}. 
However, all of these existing denoising methods perform poorly when denoising channel vectors that were acquired through low-resolution data converters. 
Channel estimation with 1-bit ADCs has been analyzed in~\cite{li17b,li16a,jacobsson17b,mollen16c,studer16a}. 
Beamspace sparsity of mmWave channels has been exploited to denoise channel vectors from 1-bit measurements in \cite{mo16b, huang19,kaushik18}.
However, all of these denoising methods exhibit high complexity, ignore beamspace sparsity, and/or require a number of parameters that must be adapted to the instantaneous propagation conditions, such as the number of dominant propagation paths.

\subsection{Contributions}
We propose low-complexity channel estimation algorithms for mmWave massive MU-MIMO systems that operate with 1-bit data converters.
By using a Bussgang-like decomposition~\cite{bussgang52a} of the 1-bit measurement process, our methods adapt the optimal denoising parameters to the channel's instantaneous sparsity via Stein's unbiased risk estimate (SURE).
We propose two methods that build upon BEACHES put forward in~\cite{ghods19a} and a novel method, referred to as Sparsity-Adaptive oNe-bit Denoiser (SAND), which automatically tunes two algorithm parameters to minimize the channel estimation mean-square error (MSE). 
To demonstrate the efficacy of our channel estimation algorithms, we perform  MSE and bit error rate (BER) simulations with  line-of-sight (LoS) and non-LoS mmWave channels in a massive MU-MIMO system. 

\subsection{Notation}
Lowercase and uppercase boldface letters denote column vectors and matrices, respectively. 
The $k$th entry of the vector~$\bma$ is~$a_k$; 
the real and imaginary parts are $\realindex{\bma} = \realpart{\bma}$ and $\imagindex{\bma} = \imagpart{\bma}$, respectively. 
For a matrix $\bA$, 
its transpose and Hermitian transpose are $\bA^\Tran$ and $\bA^\Herm$, respectively. 
A complex Gaussian vector $\bma$ with mean $\bmm$ and covariance $\bK$ is written as $\bma\sim\CN(\bmm, \bK)$.
Expectation is denoted by $\smolE{\cdot}$. 
%
\section{1-bit Quantized System Model}
\label{sec:systemmodel}

We consider a mmWave massive MU-MIMO uplink system in which $U$ single-antenna UEs transmit data to a $B$-antenna BS equipped with a uniform linear array (ULA). 
We assume that each of the $B$ radio-frequency (RF) chains at the BS contains a pair of 1-bit ADCs that separately quantize the in-phase and quadrature signals.
A widely-used, yet simplistic channel vector model for such systems is as follows \cite{tse05a}:
\begin{align} \label{eq:antenna_domain_channel}
\bmh =  \textstyle \sum_{\ell=1}^{L}{\!\kappa_\ell\bma(\Omega_\ell)}, \, \bma(\Omega)\! =\! [e^{j0\Omega}, e^{j1\Omega}, \ldots, e^{j(B-1)\Omega}]^\Tran\!\!.
\end{align}
Here, $L$ stands for the number of propagation paths arriving at the BS, $\kappa_\ell\in\complexset$ is the channel gain of the $\ell$th path, $\bma(\Omega_\ell)\in\complexset^B$ contains the relative phases between BS antennas, and $\Omega_\ell\in[0,2\pi)$ is determined by the $\ell$th path's incident angle.
We emphasize that our simulation results in \fref{sec:simulations} will use more realistic mmWave channel models. 

We consider orthogonal training-based channel estimation, where only one UE transmits a pilot at a time---a generalization to other training schemes is part of ongoing work. To model 1-bit ADCs, we define $Q(z) = \sign(\realpart{z}) + j \sign(\imagpart{z})$, which is applied element-wise to vectors. The 1-bit quantized channel vector for a given UE can be modeled as follows \cite{li17b}:
\begin{align} \label{eq:quantized_antenna_domain}
\bmr = & Q\!\left(\varrho\bmh+\bmn\right)\!,
\end{align}
where $\bmn\sim\CN(\boldsymbol0,\No\bI_B)$ models thermal noise. Without loss of generality, we assume  $\varrho=1$ for the rest of the paper. 

All of the above vectors are in the antenna domain, where each entry is associated with one of the $B$ BS antennas. 
By taking the discrete Fourier transform (DFT) across the antenna array, we can transform these vectors into the beamspace domain, where each entry corresponds to an incident angle. 
From \fref{eq:antenna_domain_channel} we see that $\bmh$ is a superposition of~$L$ complex sinusoids. Consequently, the beamspace domain representation $\hhat = \bF\bmh$, where~$\bF$ is the $B\times B$ unitary DFT matrix, will be  sparse assuming that $L\ll B$. In what follows, all beamspace domain quantities are designated with a $\hat{\text{hat}}$.

\begin{figure}[tp]
\subfigure[LoS channel]{\includegraphics[width=0.495\columnwidth]{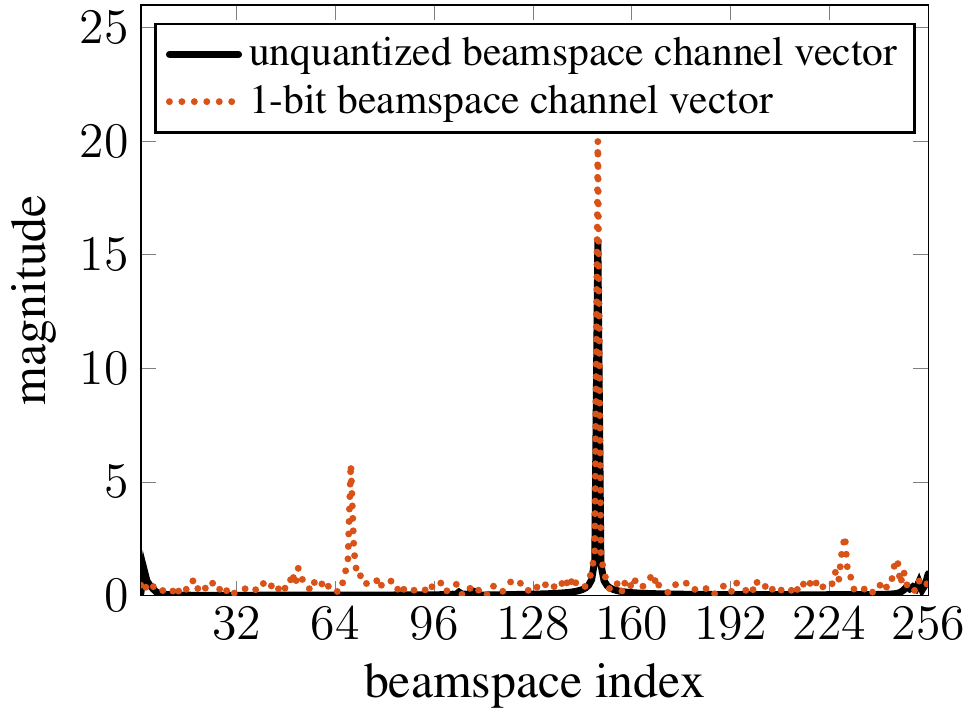}}
\subfigure[Non-LoS channel]{\includegraphics[width=0.495\columnwidth]{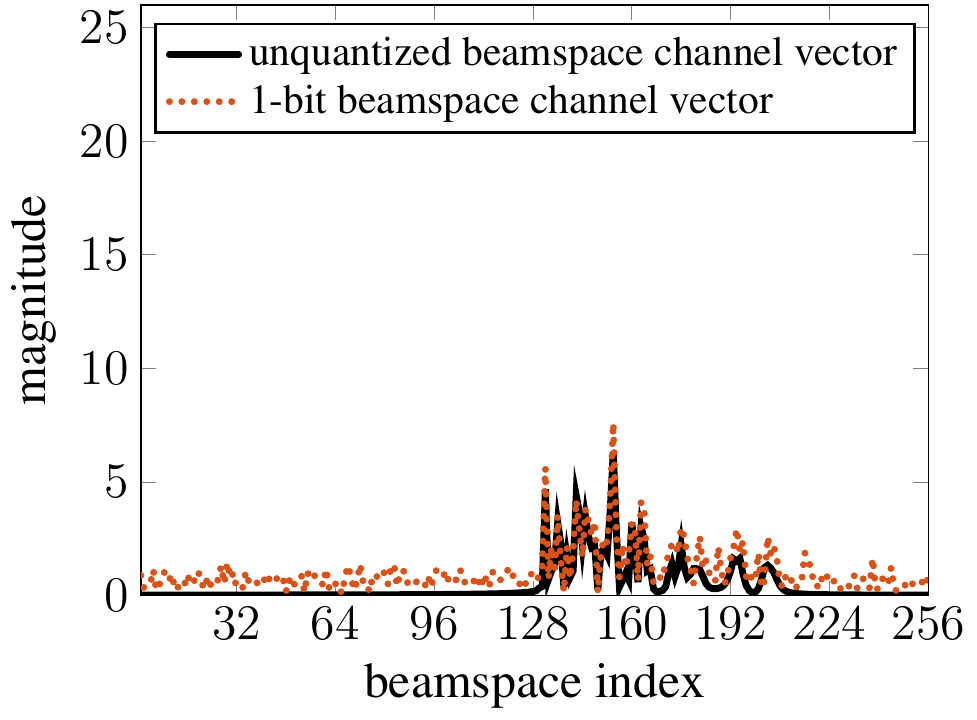}}
\vspace{-0.35cm}
\caption{Beamspace representation of an unquantized and a 1-bit quantized channel vector for a line-of-sight (LoS) and non-LoS scenario. The channel vectors are generated with the QuaDRiGa mmMAGIC UMi model \cite{jaeckel2019quadriga} at $60$\,GHz for a $256$ uniform-linear array (ULA) with $\lambda/2$ antenna spacing. The average energy has been normalized to $1$ and no noise is present.} 
\label{fig:sparse_channel}
\end{figure}

\fref{fig:sparse_channel} shows examples for LoS and non-LoS channel vectors in the beamspace domain without and with 1-bit quantization. 
Clearly, the unquantized beamspace vectors $\hhat$ exhibit sparsity; the 1-bit quantized beamspace vectors, which are obtained from $\rhat = \bF\bmr$, also exhibit sparsity but, in addition, are distorted by quantization artifacts.
We also observe that the quantization artifacts differ significantly between the LoS and non-LoS channels, which exhibit different levels of sparsity. 
In what follows, we develop algorithms that exploit beamspace sparsity to denoise 1-bit quantized channel vectors while adapting the denoising parameters to the instantaneous channel sparsity.

\section{BEACHES-Based 1-Bit Denoising}
\label{sec:one_bit_denoising}
Before we discuss denoising methods for 1-bit measurements, we briefly review the BEACHES algorithm in~\cite{ghods19a}, which was developed for systems with high-resolution data converters. Assume that we observe a noisy measurement of the channel vector $\bmh$ in the beamspace domain as
\begin{align} \label{eq:noisy_measurement}
\yhat = \hhat + \ehat,
\end{align}
where $\ehat\sim\setC\setN(\bZero,\Eo\bI)$ models channel estimation errors.
BEACHES denoises $\yhat$ by applying the soft-thresholding function $\hhattick = \eta(\yhat,\tau)$ defined as 
\begin{align}
\textstyle [\eta(\yhat,\tau)]_b = \frac{\yhate{b}}{|\yhate{b}|}\max\{|\yhate{b}|-\tau,0\},  \quad b=1,\ldots,B,
\end{align}  
where we define ${\yhate{b}}/{|\yhate{b}|} = 0$ for $\yhate{b} = 0$ and the parameter $\tau$ is the denoising threshold. 
For an optimally-chosen denoising threshold~$\tau^\star$, the soft-thresholding function suppresses noise (which is typically weak) while leaving the sparse components that pertain to the channel vector mostly intact.  We define $\tau^\star$ as the denoising threshold that minimizes the MSE  
\begin{align} \label{eq:MSE}
\textstyle \MSE = \frac{1}{B}\smolE{\|\hhattick-\hhat\|^2},
\end{align}
which determines the optimally-denoised channel vector $\hhatstar = \eta(\yhat,\tau^\star)$ in the beamspace domain. However, the MSE expression depends on the unknown vector $\hhat$. BEACHES circumvents this issue by using Stein's unbiased risk estimator (SURE) \cite{donoho95}, which is an unbiased estimate of the MSE in~\fref{eq:MSE} that does \emph{not} depend on $\hhat$. 
BEACHES requires (i) the channel estimation error $\hat\bme$ to be i.i.d.\ Gaussian
and (ii) knowledge of the channel estimation variance $\Eo$ to optimally denoise the channel vector at a complexity that scales only with $B\log(B)$.

\subsection{The $\textit{1}$-BEACHES Algorithm} \label{sec:1BEACHES}
We now present $1$-BEACHES, which denoises the received 1-bit channel measurements using BEACHES.
For this method, we model the 1-bit received vector $\bmr$  in \fref{eq:quantized_antenna_domain}   as 
\begin{align} \label{eq:1beaches_linearization}
{\bmr}=  Q(\bmh+\bmn) = \bmh + \bmq,
\end{align}
where the vector $\bmq$ depends on $\bmh$ and~$\bmn$, and models quantization errors and noise.
By transforming $\bmr$ into the beamspace domain, we have 
\begin{align} \label{eq:beamspace_domain_bussgang}
\hat{\bmr}= \bF \bmr = \hhat + \qhat.
\end{align}
Even though the vector $\bmq$ is not Gaussian distributed, the beamspace version $\qhat=\bF\bmq$ is well-approximated by a Gaussian random vector as each entry is a sum of all entries of $\bmq$ with different phases. 
To denoise the system in~\fref{eq:beamspace_domain_bussgang} with BEACHES, we need knowledge of the variance~$\Qo$ of the entries in $\qhat$.
By assuming that $\hhat=\bF\bmh$ is circularly-symmetric, which is reasonable as $\bmh$ is a sum of complex sinusoids as modeled in~\fref{eq:antenna_domain_channel}, we obtain   
\begin{align} \label{eq:expression0}
\textstyle \Qo =  \frac{1}{B}\smolE{\|\qhat\|^2} =  \frac{1}{B}\smolE{\|\bmr\|^2+\|\bmh\|^2-2\realpart{\bmh^\Herm\bmr}}.
\end{align}
In order to obtain a closed-form expression of $\Qo$ with a minimal number of parameters, we further assume\footnote{This assumption is accurate if the number of propagation paths $L$ in \fref{eq:antenna_domain_channel} is large. As shown in \fref{sec:simulations}, this assumption is simplistic for LoS channels.} that $\bmh\sim\CN(\boldsymbol0,\Eh\bI_B)$, which leads to $\Qo = 2+\Eh-{{4\Eh}/{\sqrt{\pi({\Eh}+{\No})}}}$,
where  $\Eh=\frac{1}{B}\smolE{\|\bmh\|^2}$ and~$\smolE{\bmh^\Herm\bmr}$ in~\fref{eq:expression0} is computed in  \fref{app:bussgang_gain}. 
Under these assumptions, the beamspace representation \fref{eq:beamspace_domain_bussgang} has the same form as \fref{eq:noisy_measurement}, where 
$\yhat=\rhat$ and we model $\ehat=\qhat\sim\CN(\boldsymbol0,\Qo\bI_B)$, which allows us to (i) apply BEACHES to find the optimal denoising threshold~$\tau^\star$ given~$\rhat$ and the variance~$\Qo$, and (ii) compute $\hhatstar = \eta(\rhat,\tau^\star).$ We call this procedure $1$-BEACHES.

\subsection{The $\alpha$-BEACHES Algorithm} \label{sec:alphaBEACHES}
In the model \fref{eq:1beaches_linearization}, the error $\bmq$ will be large if the power of~$\hhat$ differs from the power of $\rhat$.
We now derive $\alpha$-BEACHES which addresses this aspect.
To this end, we use a Bussgang-like decomposition \cite{bussgang52a} that models the 1-bit ADCs as 
\begin{align} \label{eq:bussgang_decomposition}
\bmr = Q(\bmh+\bmn) = \alpha \bmh + \bmd,
\end{align}
where $\alpha$ is a scalar that minimizes the distortion variance $\smolE{\|\bmd\|^2}$ and also ensures $\smolE{\bmd^\Herm\bmh}=0$.
By assuming that the vector~$\bmh$ is circularly symmetric, we have
\begin{align} \label{eq:alpha_bussgang_general}
\alpha = & 
\argmin_{\alpha^\prime\in\complexset}\smolE{\|\bmr-\alpha^\prime\bmh\|^2} 
= \textstyle \frac{\smolE{\bmh^\Herm \bmr}}{\smolE{\|\bmh\|^2}}.
\end{align}
To obtain a closed-form expression for $\alpha$, we assume $\bmh\sim\setC\setN(\bZero,\Eh\bI)$ as in 1-BEACHES
and use the derivation of $\smolE{\bmh^\Herm\bmr}$ in \fref{app:bussgang_gain},  which yields 
$\alpha = {2}/{\!\sqrt{\pi({\Eh}+{\No})}}$.

In \fref{eq:bussgang_decomposition}, the distortion $\bmd$ is not Gaussian. By transforming into beamspace domain and dividing the result by $\alpha$, we get
\begin{align} \label{eq:bussgang_decomposition_beamspace}
\textstyle \frac{1}{\alpha}\rhat = \frac{1}{\alpha}\bF \bmr =  \hhat + \frac{1}{\alpha}\dhat,
\end{align}
in which the distortion $ {\dhat}/{\alpha}$ is well-approximated by a  Gaussian, as each entry is a scaled and phase-shifted sum of all of the entries of $\bmd$. The distortion variance ${\Do}/{\alpha^2}$ is 
\begin{align}
\textstyle \frac{1}{B}\frac{1}{\alpha^2} \smolE{\|\dhat\|^2}
= \frac{1}{B}\frac{1}{\alpha^2}\smolE{\|\bmr\|^2\!-\alpha^2\|\bmh\|^2}\! = \frac{2}{\alpha^2}-E_h.
\end{align}
The model \fref{eq:bussgang_decomposition_beamspace}, enables us to apply BEACHES to $\rhat/{\alpha}$ in order to determine the denoising threshold $\tau^\star$ given $\rhat/\alpha$ and $\Do/\alpha^2$. Finally, $\alpha$-BEACHES computes $\hhatstar = \eta(\frac{\rhat}{\alpha},\tau^\star).$


\section{SAND: Sparsity-Adaptive oNe-bit Denoiser} \label{sec:SAND}

As a generalized variant of $\alpha$-BEACHES, we next develop a sparsity-adaptive method that \emph{simultaneously} learns a prefactor~$\gamma$ and a denoising threshold~$\tau$ in order to minimize the MSE.
By defining our two-parameter estimator\footnote{This estimator is equivalent to $\hhattick = \eta(\gamma'\rhat,\tau')$ for $\gamma = \gamma'$ and $\tau = \frac{\tau'}{\gamma}$.} as $\hhattick = \gamma\eta(\rhat,\tau)$, we aim to find the parameters $\gamma^\star$ and $\tau^\star$ that minimize the MSE in \fref{eq:MSE}. 
Since the MSE depends on the unknown vector~$\hhat$, we select the optimal parameters $\gamma^\star$ and $\tau^\star$ that minimize SURE, which (i) is an unbiased estimator of the MSE so that $\E{\SURE}=\MSE$ and $\lim_{B\to\infty}\SURE = \MSE$ (see \cite{ghods19a} for the details) and (ii) does \emph{not}  depend on $\hhat$. 
For any weakly differentiable estimator $\mur$, using the decomposition \fref{eq:bussgang_decomposition_beamspace} and assuming that $\dhat$ is i.i.d.\ Gaussian,  SURE is given by  
\begin{align}
\SURE = \, & \textstyle \frac{1}{B} {\|\mur\|^2}+ \frac{2-\Do}{\alpha^2}  -  \frac{1}{B} {\frac{2}{\alpha}\realpart{{\rhat}^\Herm \mur}} \notag\\
& + \textstyle \frac{1}{B} \sum_{b=1}^B \frac{\Do}{\alpha}\left(\frac{\partial\realindex{\mure{b}}}{\partial\realindex{\rhate{b}}}+\frac{\partial\imagindex{\mure{b}}}{\partial\imagindex{\rhate{b}}}\right)\!.  \label{eq:sure}
\end{align}
Refer to \fref{app:sand_sure} for the proof.
Since SURE is an unbiased estimator of the MSE, we use SURE in \fref{eq:sure} with $\mur = \gamma\eta(\rhat,\tau)$, in order to find the optimal parameters $\gamma^\star$ and $\tau^\star$.
While a na\"ive approach could perform a two-dimensional grid search over the tuple $(\gamma,\tau)$, we next show that we can efficiently find $\gamma^\star$ and $\tau^\star$ with $\setO(B\log(B))$ complexity.

Let $\rshat$ be a vector containing the absolute values of $\rhat$ sorted in ascending order. For a given $\tau$, let $k$ be the number of entries in $\rshat$ that are smaller than $\tau$.
For the denoiser $\mur = \gamma\eta(\rhat,\tau)$, following the derivations in \cite[App.~B]{ghods19a}, SURE in \fref{eq:sure} is 
\begin{align} \label{eq:SureShrink1}
\SURE = \, & \textstyle \frac{1}{B} \gamma^2\sum_{b=k+1}^{B}{(\rshate{b}-\tau)^2} + \frac{2-\Do}{\alpha^2} \\
& - \textstyle \frac{1}{B} \frac{\gamma}{\alpha}\sum_{b=k+1}^{B}\left({{2}{\rshate{b}(\rshate{b}-\tau)}}-{\Do}\left(2-\textstyle\frac{\tau}{\rshate{b}}\right)\right) \notag.
\end{align}
By defining the quantities $\sumabsysquared = \sum_{b=k+1}^{B} (\rshate{b})^2$, $\sumabsy = \sum_{b=k+1}^{B} \rshate{b}$ and $\sumabsyinverse = \sum_{b=k+1}^{B} (\rshate{b})^{-1}$, we can rewrite \fref{eq:SureShrink1} as
\begin{align} 
\SURE = \, & \textstyle \frac{1}{B}\gamma^2\left(\sumabsysquared-2\tau\sumabsy+\tau^2\sumone\right) + \frac{2-\Do}{\alpha^2} \notag \\
& - \textstyle \frac{1}{B}\frac{\gamma}{\alpha}\left(2\left(\sumabsysquared-\tau\sumabsy\right)  
- \Do\left(2\sumone-\tau\sumabsyinverse\right)\right)\!.
\label{eq:SureShrinkabc}
\end{align}
For a fixed $\tau$, the optimal $\gamma^\star\in\reals_{\geq0}$ that minimizes \fref{eq:SureShrinkabc} is
\begin{align} \label{eq:gammastar}
\gamma^\star = \max\{0,\textstyle \frac{2\left(\sumabsysquared-\tau\sumabsy\right)  
	- \Do\left(2\sumone-\tau\sumabsyinverse\right)}{2\alpha\left(\sumabsysquared-2\tau\sumabsy+\tau^2\sumone\right)}\}.
\end{align}
The optimal threshold $\tau^\star$ could take any value between $0$ and~$\rshate{B}$. However, as in the derivation of BEACHES~\cite{mirfarshbafan19a}, we restrict the search to values in $\rshat$, as it significantly reduces the complexity, without sacrificing performance. We also set an upper limit for $\tau$ of $\sqrt{2\Do\log(B)}$, which ensures (with high probability) that the threshold is lower than the largest noise realization~\cite{donoho95}.
For each $\tau = \rshate{k}$, $k = 0,\ldots,B$ (with $\rshate{0} = 0$), and for its associated $\gamma^\star$ given by \fref{eq:gammastar}, we evaluate SURE in \fref{eq:SureShrinkabc}, and then pick $\gamma^\star$ and $\tau^\star$ that result in the minimum value of SURE. We call the resulting algorithm Sparsity-Adaptive oNe-bit Denoiser (SAND), which is summarized in \fref{alg:SAND}. 
Since the complexity of a fast Fourier transform (FFT) and sorting scale with $\mathcal{O}(B\log(B))$, and the operations in each iteration (lines 6 to 11) have complexity $\setO(1)$, the overall complexity of SAND scales with $\mathcal{O}(B\log(B))$.

\makeatletter
\newcommand\fs@betterruled{%
  \def\@fs@cfont{\bfseries}\let\@fs@capt\floatc@ruled
  \def\@fs@pre{\vspace*{5pt}\hrule height.8pt depth0pt \kern2pt}%
  \def\@fs@post{\kern2pt\hrule\relax}%
  \def\@fs@mid{\kern2pt\hrule\kern2pt}%
  \let\@fs@iftopcapt\iftrue}
\floatstyle{betterruled}
\restylefloat{algorithm}
\makeatother
\setlength{\textfloatsep}{5pt}
\begin{algorithm}[tp]
\caption{\strut SAND: Sparsity-Adaptive oNe-bit Denoiser \label{alg:SAND}}
\begin{algorithmic}[1]
\STATE {\bf input} $\bmr$, $\alpha$ and $\Do$ 
\STATE $\rhat = \text{FFT}(\bmr)$, $\SURE_{\text{min}}=\infty$	, $\tau=0$			
\STATE $\rshat=\text{sort}\{|\rhat|,\text{`ascend'}\}$, $\rshate{B+1}=\rshate{B+2}=\infty$
\STATE $\sumabsysquared=\sum_{k=1}^{B} {(\rshate{k})^{2}}$, $\sumabsy=\sum_{k=1}^{B} {\rshate{k}}$, $\sumabsyinverse=\sum_{k=1}^{B} {(\rshate{k})^{-1}}$
\FOR{$k=0,\ldots,B+1$}
\STATE $\gamma =\max\{0, \frac{2(\sumabsysquared-\tau \sumabsy) - \Do(2\sumone-\tau \sumabsyinverse)}{2\alpha(\sumabsysquared-2\tau \sumabsy+\tau^2 \sumone)}\}$
\STATE $\SURE = \frac{1}{B}\gamma^2\left(\sumabsysquared-2\tau\sumabsy+\tau^2\sumone\right) + \frac{2-\Do}{\alpha^2}
- \frac{1}{B}\frac{\gamma}{\alpha}\left(2\left(\sumabsysquared-\tau\sumabsy\right) 
- \Do\left(2\sumone-\tau\sumabsyinverse\right)\right)$ 					
\IF{$\SURE<\SURE_{\text{min}}$ \AND $\tau<\sqrt{2\Do\log(B)}$}		
\STATE $\SURE_{\text{min}} = \SURE$, $\tau^\star = \tau$, $\gamma^\star = \gamma$
\ENDIF
\STATE $\tau \!=\! \rshate{k+1}$, $\sumabsysquared \!=\! \sumabsysquared \!-\! (\rshate{k+1})^2, \sumabsy \!=\! \sumabsy \!-\! \rshate{k+1}, \sumabsyinverse \!=\! \sumabsyinverse \!-\! (\rshate{k+1})^{-1}$
\ENDFOR																				
\STATE $\hhatstare{k}=\gamma^\star\frac{\rhate{k}}{|\rhate{k}|}\max{\{|\rhate{k}|-\tau^\star,0\}}$, $k=1,\ldots,B$
\STATE  {\bf return} $\bmh^\star = \text{IFFT}(\hhatstar)$ 
\end{algorithmic}
\end{algorithm}

\section{Results}
\label{sec:simulations}

We now demonstrate the efficacy of 1-BEACHES, $\alpha$-BEACHES, and SAND. 
As reference methods, we consider perfect channel state information (CSI), where $\bmh^\star= \bmh$,  BEACHES \cite{ghods19a}, which denoises the infinite-resolution (unquantized) measurements $\bmy = \bmh+\bmn$, and 1-bit maximum-likelihood (ML) channel estimation, where $\bmh^\star = \bmr$ is the 1-bit observation in~\fref{eq:quantized_antenna_domain}. In addition, we compare the performance to state-of-the-art denoising methods, including (i) Newtonized orthogonal matching pursuit (NOMP) \cite{mamandipoor16} with an equivalent noise variance $\Qo$  and a false alarm rate $P_\text{fa}=0.5$ (using $\Eo$ results in poor performance; $P_\text{fa}$ has been tuned to achieve low MSE at low and high SNR)
and (ii) the 1-bit Bussgang linear MMSE  estimator (BLMMSE)~\cite{li16a,li17b}, which corresponds to $\bmh^\star = \frac{\Eh}{\sqrt{\pi(\Eh+\No)}}\bmr$ for the used orthogonal pilots.

\begin{figure}[tp]
	\centering
	\subfigure[LoS channel]{\includegraphics[width=0.235\textwidth]{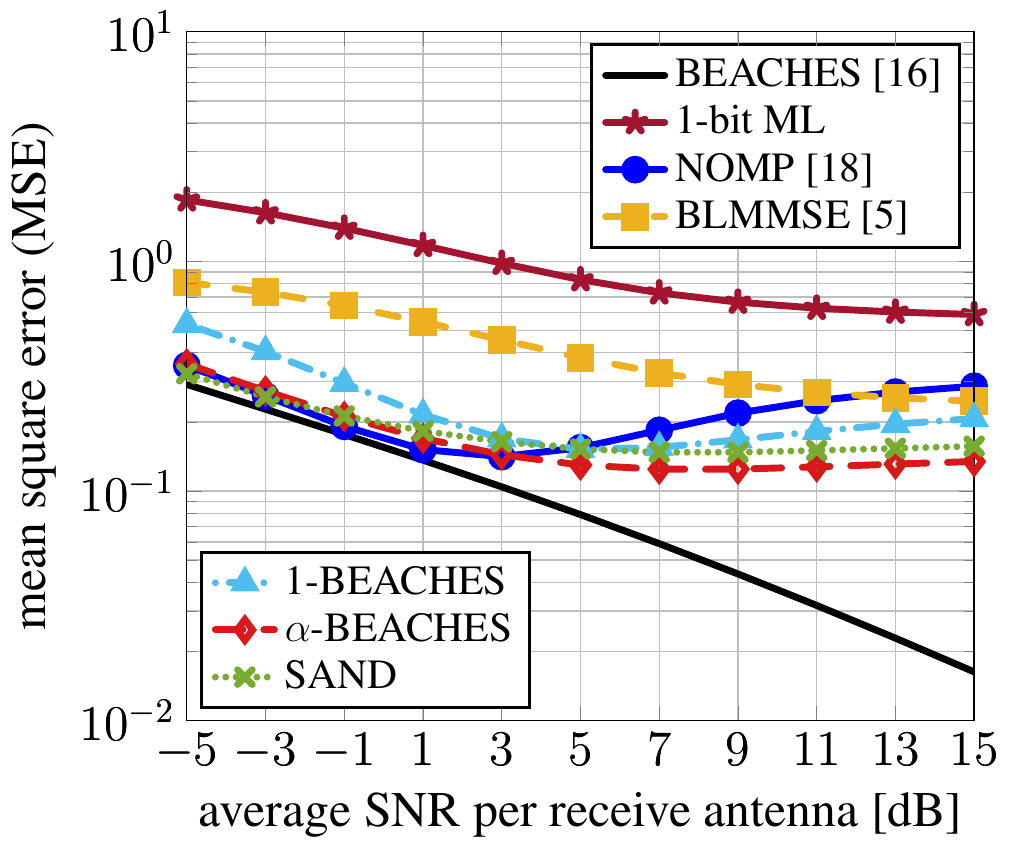}}
	\subfigure[non-LoS channel]{\includegraphics[width=0.235\textwidth]{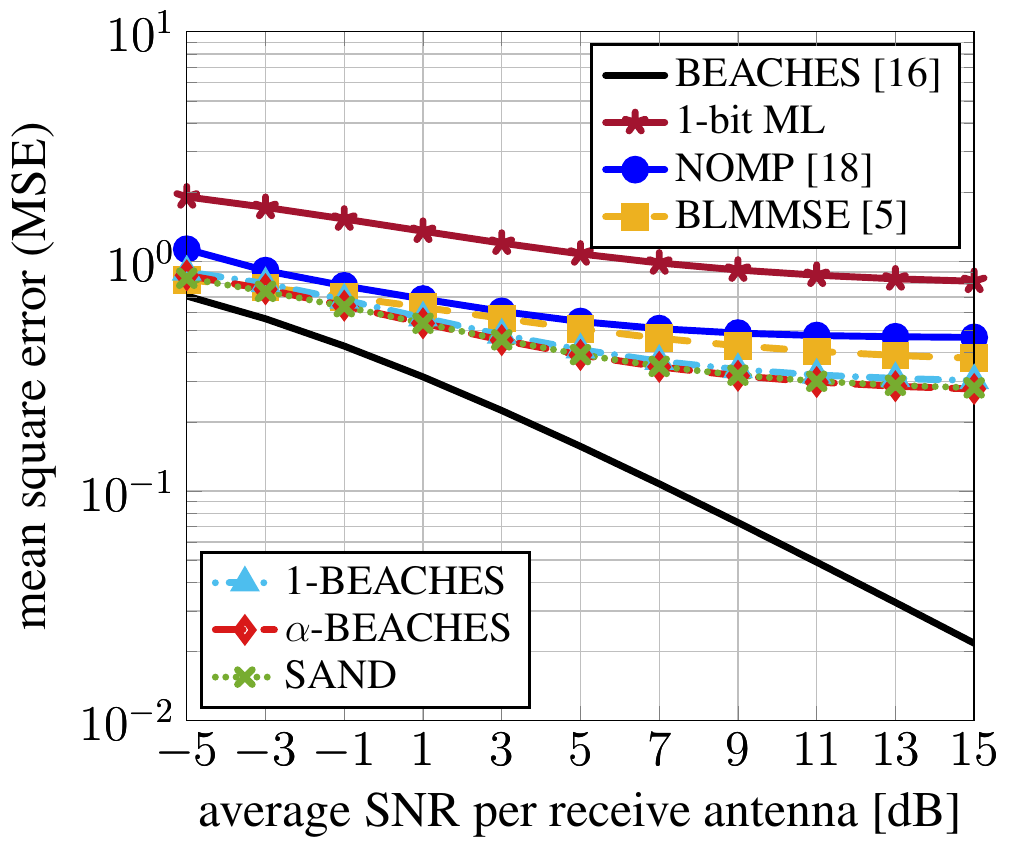}}
	\vspace{-0.1cm}
	\caption{Mean square error (MSE) of the considered channel denoising methods for mmWave LoS and non-LoS channels. The proposed sparsity-adaptive denoising methods significantly outperform na\"ive 1-bit ML channel estimation.}
	\label{fig:MSE}
\end{figure}

\subsection{Simulation Setup}
We simulate a mmWave massive MIMO system with $B=256$ BS antennas and $U=16$ single-antenna UEs. 
We generate LoS and non-LoS channel matrices using the QuaDRiGa mmMAGIC UMi model \cite{jaeckel2019quadriga} at a carrier frequency of $60$\,GHz for a BS with $\lambda/2$-spaced antennas arranged in a ULA. 
The UEs are placed randomly in a $120^\circ$ circular sector around the BS between a distance of $10$\,m and $110$\,m, and the UEs are separated by at least $4^\circ$. 
We model UE-side power control to ensure that the highest receive power is at most $6$\,dB higher than that of the weakest UE.

\begin{figure*}[tp]
	\centering
	\subfigure[LoS, QPSK]{\includegraphics[width=0.24\textwidth]{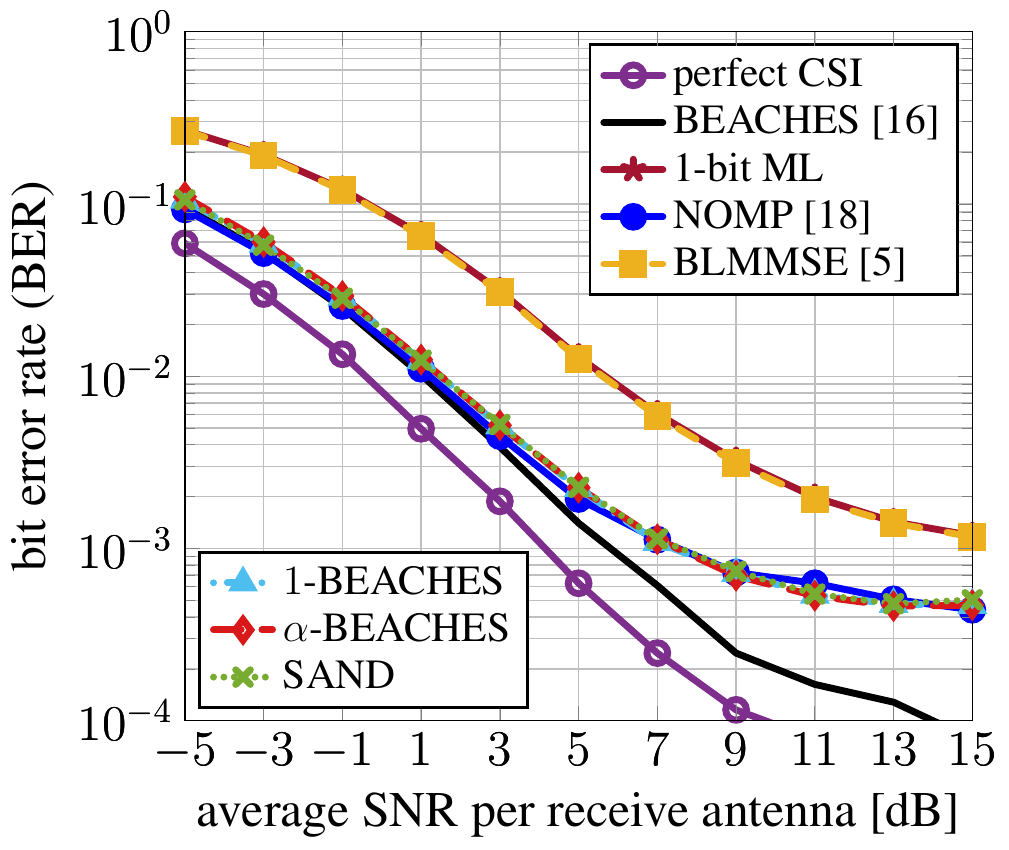}}
	\hfill
	\subfigure[non-LoS, QPSK]{\includegraphics[width=0.24\textwidth]{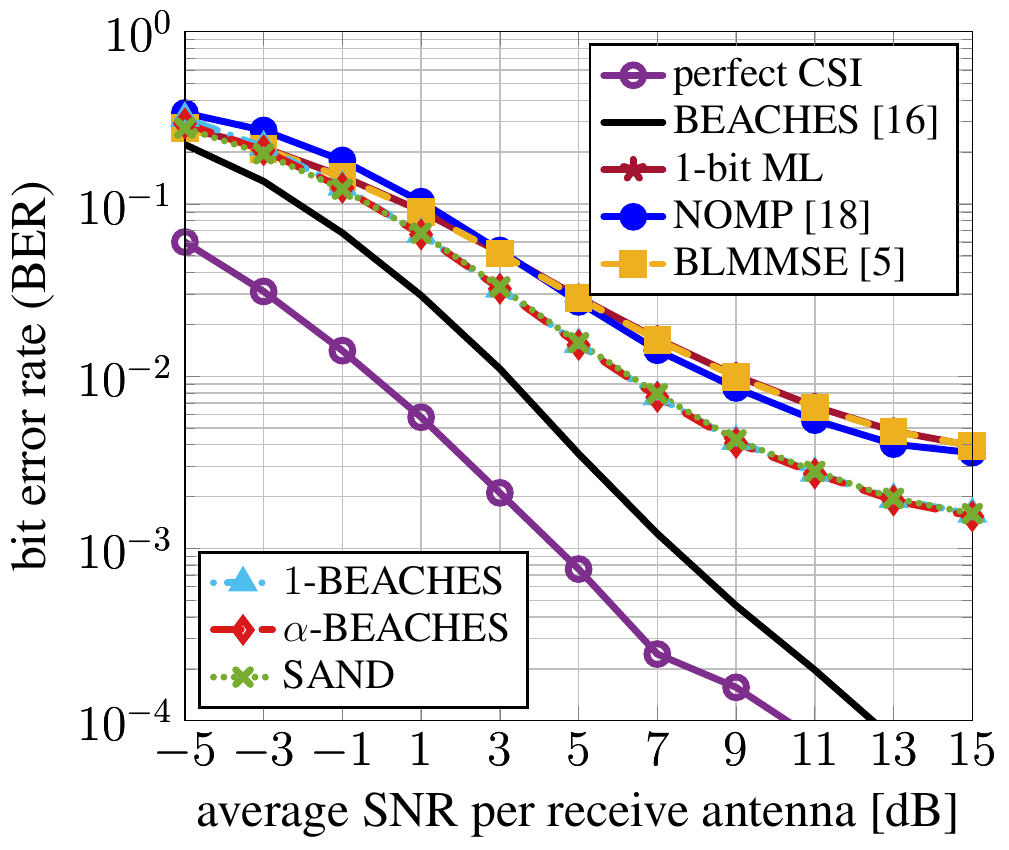}}
	\hfill
	\subfigure[LoS, 16-QAM]{\includegraphics[width=0.24\textwidth]{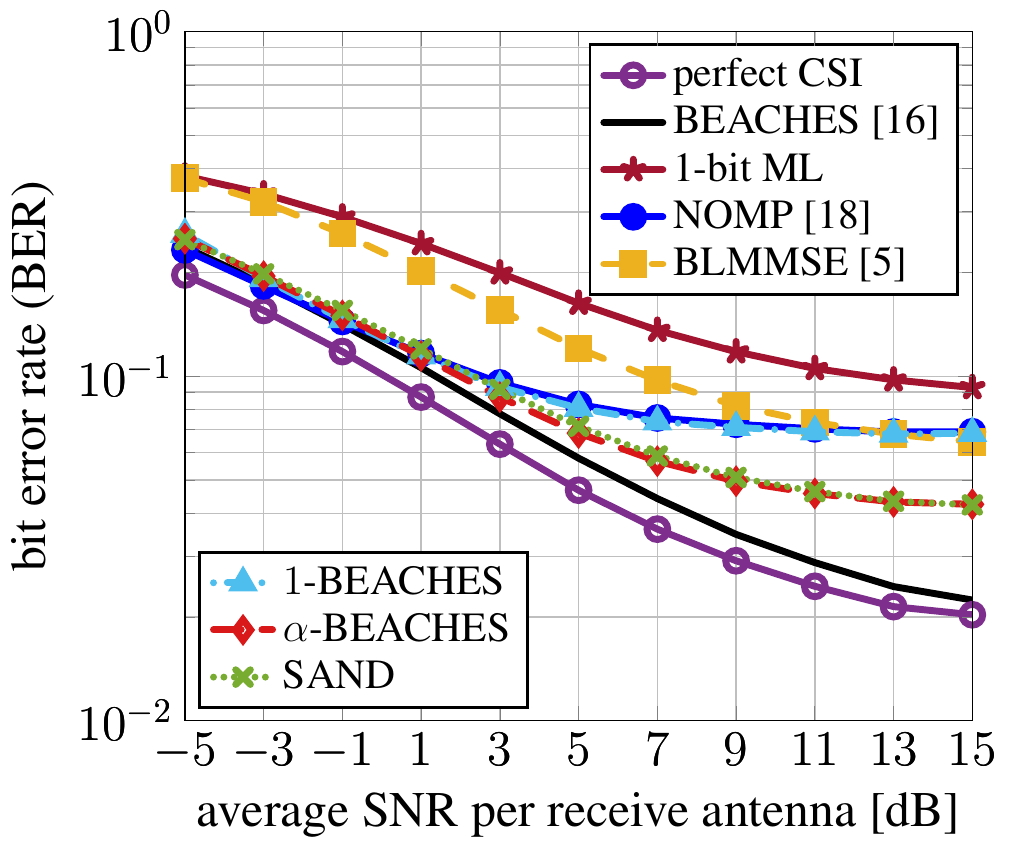}}
	\hfill
	\subfigure[non-LoS, 16-QAM]{\includegraphics[width=0.24\textwidth]{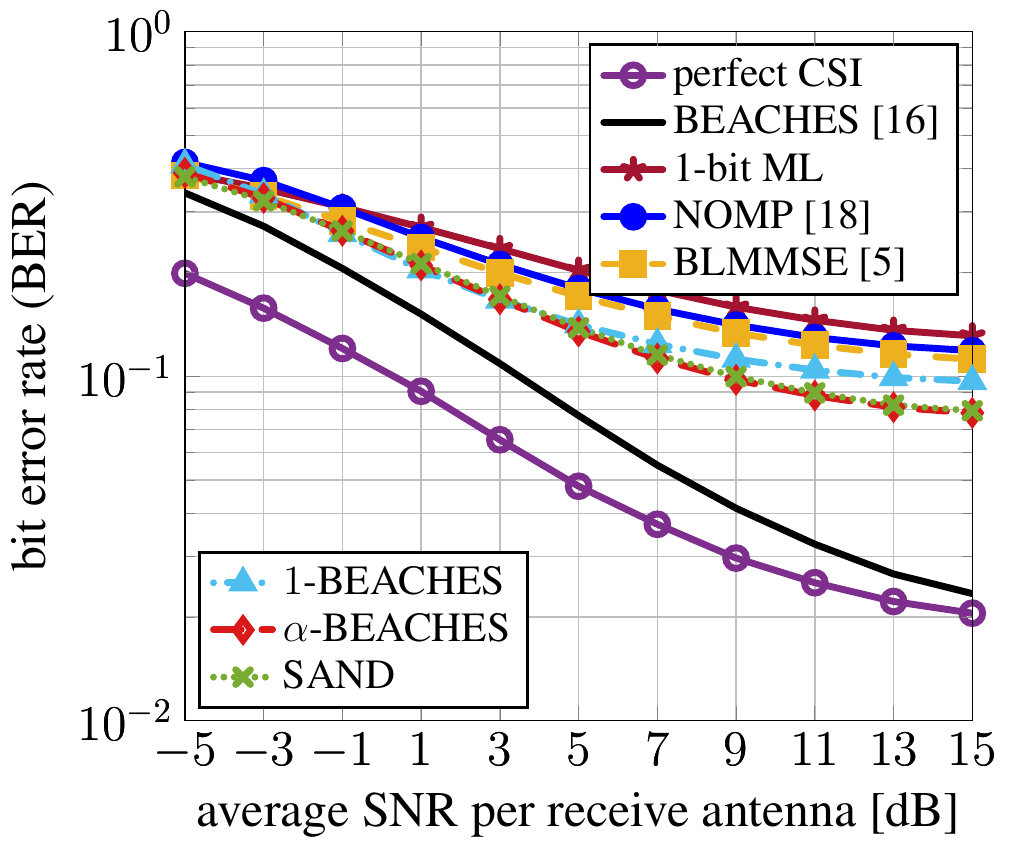}}
	\vspace{-0.1cm}
	\caption{Uncoded bit error rate (BER) of 1-bit channel estimation and 1-bit data detection in mmWave LoS and non-LoS channels. We see that $\alpha$-BEACHES and SAND outperform $1$-BEACHES and 1-bit ML for LoS and non-LoS channel conditions for 16-QAM transmission. }
	\label{fig:BER}
\vspace{-0.37cm}	
\end{figure*}

\subsection{Mean-Square Error (MSE) Performance}
\fref{fig:MSE} shows the channel estimation MSE of the proposed 1-bit denoising algorithms and the considered baseline methods. We observe that the three proposed methods, $1$-BEACHES, $\alpha$-BEACHES, and SAND significantly outperform 1-bit ML channel estimation. Furthermore, we see that $\alpha$-BEACHES and SAND have a slight advantage over $1$-BEACHES in LoS scenarios. Surprisingly, SAND has a slightly higher MSE than $\alpha$-BEACHES, which we attribute to the fact that SAND has to learn two parameters, whereas $ \alpha$-BEACHES only learns the optimal denoising threshold. For that reason, SAND is more sensitive to the assumptions made in footnote~1.
NOMP and BLMMSE also outperform 1-bit ML, but their MSE is higher than that of our algorithms, especially at high SNR. 

\subsection{Bit Error Rate (BER) Performance}
To assess the impact of the proposed 1-bit denoising algorithms on the uncoded BER performance during the data detection phase, we use the 1-bit Bussgang linear MMSE equalizer proposed in \cite{nguyen19}, which operates on the 1-bit quantized received data using the channel estimates provided by our denoising methods and the considered baseline algorithms. We consider QPSK and $16$-QAM transmission.

\fref{fig:BER} shows that the proposed sparsity-adaptive denoising algorithms significantly outperform na\"ive 1-bit ML channel estimation. We furthermore see that for QPSK, all three methods,  $1$-BEACHES, $\alpha$-BEACHES, and SAND, perform equally well under both LoS and non-LoS scenarios. For 16-QAM, where it is important to get an accurate estimate of the channel gain, $\alpha$-BEACHES and SAND outperform $1$-BEACHES and NOMP, which directly operate with the received 1-bit measurements. 
Hence, correcting the scale of the received data is critical for higher-order constellation sets.
While BLMMSE adjusts for the scale, it is unable to exploit sparsity which results in rather poor BER performance.
For non-LoS channels, NOMP performs inferior to the proposed methods. 
In addition, NOMP requires high complexity~\cite{mirfarshbafan19a}.
Since the propagation conditions (such as the number of propagation paths $L$) are typically unknown in practice, SAND and $\alpha$-BEACHES are the preferred denoising methods. 
%


\section{Conclusions}
\label{sec:conclusions}

We have presented three sparsity-adaptive channel vector denoising algorithms for 1-bit mmWave massive MIMO systems.
Two of our algorithms denoise 1-bit measurements of the channel estimates using BEACHES~\cite{ghods19a} in order to automatically adapt the denoising parameter to the instantaneous channel realization. 
While 1-BEACHES applies  BEACHES to the 1-bit measurements using the effective noise variance (which also includes the quantization noise variance), $\alpha$-BEACHES uses a Bussgang-like scaling factor~\cite{bussgang52a}, which results in superior performance.
We have also introduced SAND (short for Sparsity-Adaptive oNe-bit Denoiser), a novel denoising algorithm with $\setO(B\log(B))$ complexity, which jointly optimizes the thresholding parameter and the scaling factor in a nonparametric fashion. 
Our simulations have shown that $\alpha$-BEACHES and SAND perform equally well under the considered LoS and non-LoS mmWave channels and outperform $1$-BEACHES as well as other considered baseline methods in the case of 16-QAM transmission.
\appendices 

\refstepcounter{section} \label{app:bussgang_gain}
\section*{\fref{app:bussgang_gain}: Derivation of $\frac{1}{B}\E{\bmh^\Herm\bmr}$} 
Since $\bmn\sim\CN(\boldsymbol0,\No\bI_B)$ and $\bmh$ is assumed circularly symmetric, the imaginary part of $\E{\bmh^\Herm\bmr}$ is zero, and
\begin{align}
\textstyle\frac{1}{B}\E{\bmh^\Herm\bmr} = 
\frac{1}{B}\E{\realindex{\bmh}\realindex{\bmr}} + \frac{1}{B}\E{\imagindex{\bmh}\imagindex{\bmr}}\!.
\end{align}
By assuming $\bmh\sim\CN(\boldsymbol0,\Eh\bI_B)$, $\frac{1}{B} \E{\realindex{\bmh}\realindex{\bmr}}$ becomes
\begin{align}
\textstyle\frac{1}{B} & \textstyle\sum_{b=1}^{B}\E{\int_{-\infty}^{-\realindex{n_b}}\frac{-\realindex{h_b}}{\sqrt{\pi\Eh}}e^{-\frac{(\realindex{h_b})^2}{\Eh}}d\realindex{h_b}} \\
\textstyle + & \textstyle\frac{1}{B}\sum_{b=1}^{B}\E{\int_{-\realindex{n_b}}^{-\infty}\!\frac{\realindex{h_b}}{\sqrt{\pi\Eh}}e^{-\frac{(\realindex{h_b})^2}{\Eh}}d\realindex{h_b}}
\!=\! \textstyle\frac{\Eh}{\sqrt{\pi(\Eh+\No)}}. \notag 
\end{align}
Following the same procedure for the imaginary part, we get
\begin{align}
\textstyle\frac{1}{B}\E{\bmh^\Herm\bmr} = \textstyle\frac{2\Eh}{\sqrt{\pi({\Eh}+{\No})}}.
\end{align}
\vspace{-0.4cm}

\refstepcounter{section} \label{app:sand_sure}
\section*{\fref{app:sand_sure}: Derivation of SURE in \fref{eq:sure}} 

For deriving SURE as in \fref{eq:sure}, we follow the procedure in \cite[App.~A]{ghods19a}, with the following modifications: Instead of $\yhat = \hhat+\ehat$, we use $\rhat = \alpha\hhat + \dhat$. In other words, where~\cite{ghods19a} uses $\yhat\sim\CN(\hhat,\Eo\bI_B)$, we replace it by $\rhat\sim\CN(\alpha\hhat,\Do\bI_B)$. Instead of $g(\yhat) = \mu(\yhat) - \yhat$, we use $g(\rhat) = \mu(\rhat) - \rhat/\alpha$.

\balance
\bibliographystyle{IEEEtran}
\bibliography{bib/confs-jrnls,bib/IEEEabrv,bib/publishers,bib/vipbib,bib/sm_ref}
\balance

\end{document}